\documentclass[journal]{IEEEtran}
\ifCLASSINFOpdf
\else
\fi
\usepackage{amsfonts}
\usepackage{amsmath,amsfonts,amssymb,epsfig,subfigure,cite,url,multirow,booktabs}
\usepackage{afterpage,mathrsfs,algorithm,algorithmic}
\usepackage{amsthm,multirow,makecell,array}

\renewcommand\arraystretch{1.5}

\begin{document}
%
\title{When Physical Layer Key Generation Meets RIS: Opportunities, Challenges, and Road Ahead}
%
%
%


\author{Ning Gao,~\IEEEmembership{Member,~IEEE,}~Yu Han,~\IEEEmembership{Member,~IEEE,}~Nannan Li,~Shi Jin,~\IEEEmembership{Senior Member,~IEEE,}\\and~Michail Matthaiou,~\IEEEmembership{Fellow,~IEEE}

\thanks{N. Gao and N. Li are with the School of Cyber Science
and Engineering, Southeast University, Nanjing 210096, China (e-mail:
ninggao@seu.edu.cn; linannan@seu.edu.cn).}
\thanks{Y. Han and S. Jin are with the National Mobile Communications
Research Laboratory, Southeast University, Nanjing 210096, China, (e-mail: hanyu@seu.edu.cn; jinshi@seu.edu.cn).}
\thanks{
M. Matthaiou is with the Centre for Wireless Innovation (CWI), Queen’s University Belfast, Belfast BT3 9DT, U.K. (e-mail: m.matthaiou@qub.ac.uk).}
}

%
%

\markboth{}%
{Shell \MakeLowercase{\textit{et al.}}: Bare Demo of IEEEtran.cls for IEEE Journals}
%



\maketitle

\begin{abstract}
Physical layer key generation (PLKG) is a promising technology to obtain symmetric keys between a pair of wireless communication users in a plug-and-play manner. The shared entropy source almost entirely comes from the intrinsic randomness of the radio channel, which is highly dependent on the wireless environments. However, in some static/block fading wireless environments, the intrinsic randomness of the wireless channel is hard to be guaranteed. Very recently, thanks to reconfigurable intelligent surfaces (RISs) with their excellent ability on electromagnetic wave control, the wireless channel environment can be customized. In this article, we overview the RIS-aided PLKG in static indoor environments, including its channel model and hardware architectures. Then, we propose potential application scenarios and analyze the design challenges of RIS-aided PLKG, including channel reciprocity, RIS reconfiguration speed and RIS deployment via proof-of-concept experiments on a RIS-aided PLKG prototype system. In particular, our experimental results show that the key generation rate is 15-fold higher than that without RIS in a static indoor environment. Next, we design a RIS jamming attack via a prototype experiment and discuss its possible attack-defense countermeasures. Finally, several conclusions and future directions are identified.
\end{abstract}

\begin{IEEEkeywords}
Endogenous security, physical layer key generation, reconfigurable intelligent surface, 6G.
\end{IEEEkeywords}

%
\IEEEpeerreviewmaketitle
\section{Introduction}
From the fifth-generation (5G) wireless communication to the forthcoming 6G wireless communication, we are progressing towards the era of Internet of Everything (IoE) with great momentum. This transformative shift is attributed to massive multiple-input multiple-output (MIMO), millimeter wave (mmWave) communication, integrated space-to-ground communication, and so on. However, due to the broadcast nature of wireless networks, malicious users can easily launch a series of attacks through the physical layer, such as jamming, eavesdropping and media access control (MAC) spoofing, etc \cite{7539590}. As more and more ubiquitous wireless networks are rolled out, the investigation of the lightweight and low latency physical layer security (PLS) becomes more important. Thus, integrating security into the physical layer is indispensable for the evolution of wireless communications. Traditionally, symmetric encryption schemes play an important role in information security, such as providing information confidentiality, information integrity and authentication. On the other hand, the secret keys management for tremendous heterogeneous Internet of Things (IoT) devices, including key generation, updates, and storage, is constantly under significant pressure.

Physical layer key generation (PLKG) is a promising technology to extract symmetric keys from wireless
fading channel in a plug-and-play manner \cite{8883129}. Specifically, the PLKG is based on short-term channel reciprocity, spatial channel uniqueness and  intrinsic channel randomness, which require no public key infrastructure (PKI). From the perspective of information-theoretical security, PLKG stands out as one of the most promising scheme for achieving Shannon's perfect encryption. The standard process of PLKG can be described as follows:
\begin{itemize}
  \item \textit{Channel probing}: Based on the short-term channel reciprocity in time division duplex (TDD) systems, the legitimate users transmit their pilot sequences accordingly to estimate the channel and collect the channel probing characteristics, such as the received signal strength (RSS) and channel state information (CSI).
  \item \textit{Quantization}: The legitimate users independently quantify the channel features into binary bit sequences, which are used as raw bit sequences. Due to the quantization accuracy, noise and imperfect synchronization, etc, there are some mismatched bits in the raw sequences.
  \item \textit{Information reconciliation}: The legitimate users negotiate the possible bit disagreements between each other by using an error correcting code, i.e., low density parity check (LDPC) code or principle component analysis, etc. Then, we obtain the raw key sequences.
  \item \textit{Privacy amplification}: To remove the possible information leakage in public negotiation, the final symmetric key is distilled from the discussed raw key sequences via the hash function, which completes the PLKG process.
\end{itemize}

However, the performance of the PLKG is strongly dependent on the intrinsic channel randomness. The key generation rate cannot be guaranteed in harsh wireless environments, yet, the data throughput is on the order of Gbit/s, which limits its practical large-scale penetration and deployment. For example, in static indoor environments, negotiating a sufficiently random raw key is a laborious and time-consuming task due to the fact that the channel based attenuations are almost predictable. This situation is predominant in some scenarios, such as inside empty rooms or at corridors during night. Previous works have focused on PLKG in harsh wireless environments \cite{6311224,9000831,1528749}. The widely studied approach is to increase the randomness of the wireless channel by employing a single relay and/or cooperative relays \cite{6311224}. The participation of untrusted relays can cause an information leakage for the secret key, and the deployment of additional trusted relays can increase the cost of PLKG. Artificial random source assistance is alternative method to improve the key generation rate \cite{9000831}. Although this method utilizes not only the channel intrinsic randomness but also the signal randomness, it needs to modify the frame structure of the pilot signal, which limits its application on existing commercial devices, i.e., Wi-Fi. What is more, intelligent antennas have been studied to provide a high fluctuation of the wireless channel, thereby extracting the high-entropy secret key \cite{1528749}. Nevertheless, the scalability and compatibility of off-the-shelf devices are hindrances to practical applications.

Reconfigurable intelligent surfaces (RISs), with their excellent ability on electromagnetic wave control, can customize the wireless channel to change the radio endogenous environment with low cost and low energy consumption, and for these reasons are becoming a potential innovative technology for 6G PLS \cite{9326394}. At present, with this excellent ability, RISs are gradually coming at the research forefront for assisting PLKG. As an early attempt, a programmable metasurface, namely HyperSurface, has been developed to show groundbreaking performance and security potential in indoor wireless communication \cite{LIASKOS20191}. Moreover, the RIS units optimization and the prototype system measurement of RIS-aided PLKG have been respectively studied in recent works \cite{9298937,9569556,9771319}. On the other hand, from an attack perspective, RIS based attacks for PLKG have been investigated, such as environment reconstruction attack and RIS manipulating attack, which make the defense strategy of PLKG even more challenging \cite{8957701}. However, we highlight that the measurements of the actual performance of the RIS-aided PLKG are not enough, while the practical design challenges associated with this technology remain unknown. Therefore, the real-world performance and design challenges of RIS-aided PLKG in static indoor environments should be further studied. This discussion is the motivating factor of this article in the filed of RIS-aided PLKG. We start our analysis with the RIS-aided channel model and hardware architecture; then, we heuristically present potential application scenarios for RIS-aided PLKG and present proof-of-concept experiments using a prototype system to discuss the design challenges, including channel reciprocity, RIS reconfiguration speed and RIS deployment. Moreover, we design a RIS jamming attack and discuss its feasible attack-defense countermeasures. Some insightful conclusions and future directions are identified.

\section{System Model}
In this section, we first provide the RIS-aided channel model, and then give the hardware architecture of the RIS.
\subsection{Channel Model}
We consider a Alice-Bob-Eve network in a static indoor environment, where transmitter Alice and Bob are the legitimate users and Eve is a malicious user. All participants are equipped with a single antenna and work in the TDD mode. Alice and Bob plan to generate a common secret key from the wireless fading channel, whilst Eve plans to hear the PLKG information over the wireless fading channel. When the RIS is added to the Alice-Bob-Eve network, the RIS-aided channel can be roughly written as the sum of the multiplicative channel and the direct link channel. Thus, the RIS-aided channel from Alice to Bob can be denoted as
\begin{equation}
\label{A2B}
\widetilde{h}_{AB}=\underbrace{\mathbf{h}^T_{RB}\mathbf{\Phi}\mathbf{h}_{AR}}_{\text{Multiplicative
channel}}+\underbrace{h_{AB}}_{\text{Direct link channel}},
\end{equation}
and the RIS-aided channel from Bob to Alice is given by
\begin{equation}
\label{A2B}
\widetilde{h}_{BA}=\mathbf{h}^T_{RA}\mathbf{\Phi}\mathbf{h}_{BR}
+h_{BA},
\end{equation}
where $\mathbf{h}_{AR}$ and $\mathbf{h}_{BR}$ represent the channel from Alice to RIS and from Bob to RIS, $\mathbf{h}_{RA}$ and $\mathbf{h}_{RB}$ represent the channel from RIS to Alice and from RIS to Bob, $(\cdot)^T$ is the transpose operation, and $\mathbf{\Phi}$ denotes the reflection matrix of the RIS in bidirectional channel probing, respectively. Similarly, the channel from Alice to Eve or from Bob to Eve can be written-out in the same manner by substituting $\mathbf{h}^T_{RE}\mathbf{\Phi}\mathbf{h}_{AR}$ or $\mathbf{h}^T_{RE}\mathbf{\Phi}\mathbf{h}_{BR}$ for the multiplicative channel and substituting $h_{AE}$ or $h_{BE}$ for the direct link channel.
\subsection{Hardware Architecture}
A RIS is made of a planar digitally programmable metasurfaces \cite{CuiTieJunRIS}. Specifically, a RIS is typically composed of three layers and a smart controller. The outer layer contains a large number of periodically repeated metasurface units, which can act directly on the incident electromagnetic signals. The subwavelength metamaterial units are composed of individual units equivalent to ``molecules/atoms" of natural materials. The middle layer is a metal isolation plate, which is used to avoid electromagnetic leakage. The inner layer is the control circuit, which is used to adjust the reflection amplitude and/or phase shift of each metasurface unit. The smart controller of the RIS is usually a programmable field-programmable gate array (FPGA), which can send coding sequence to the RIS and connect wirelessly to communication components, i.e., access points and terminals. In this case, by using FPGA as a smart controller, the RIS can realize different wireless propagation functions. Theoretically, the reflection amplitude and/or phase can be continuously adjusted, where the reflection amplitude can be effectively customized within $[0,1]$ by changing the resistor load, while the reflection phase can be shifted within $[0,\pi]$ by designing binary coding sequences. However, due to the hardware cost and implementation complexity, the existing works often consider the discrete control with finite amplitude and/or phase values. Take $1$-bit phase shift control as an example: each metasurface unit can independently realize 0 and $\pi$ phase shifts by switching the PIN diode between ``OFF" and ``ON" states, respectively.
\section{Potential Application Scenarios}
\begin{figure*}[!ht]
 \centering
  \includegraphics[width=16.2cm]{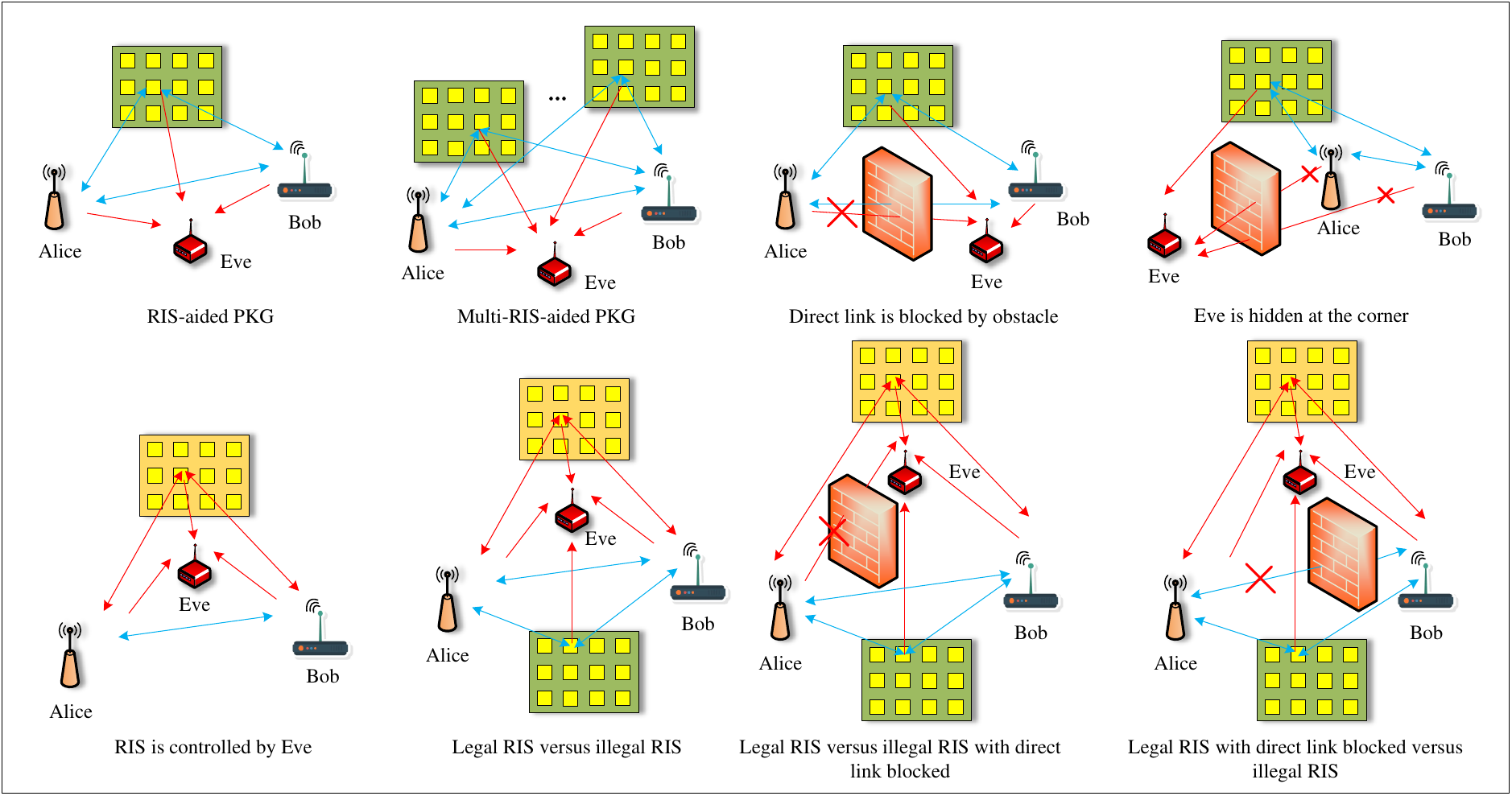}
  \caption{The potential application scenarios for RIS-aided PLKG.}
  \label{Fig:RIS_scenario}
\end{figure*}

In the past five years, several works have been reported on RIS-aided PLS for extremely diversified scenarios, i.e., RIS in unmanned aerial vehicle (UAV) secure communications \cite{9656117}. A brief summary is given in Table \ref{Tab:scenarios} and more details can be found in \cite{9198898}.
\begin{table*}[h]
\renewcommand\arraystretch{1.33}
\caption{Short summary of RIS-aided keyless PLS scenarios.}
\label{Tab:scenarios}
\centering
\begin{tabular}{c|c|c}
  \hline
  \textbf{RIS state} & \textbf{Common scenario} & \textbf{Metric} \\
  \hline
  Stationary RIS& RIS on the facade of buildings/indoor walls& \multirow{3}*{\makecell[c]{\textbf{Keyless PLS:} Secrecy capacity (SC)/rate (SR)/outage probability (SOP)/\\outage capacity (SOC)/energy efficiency,
  average SC/SR/SOP/SOC.\\ Average secrecy outage rate/duration, amount of secrecy loss.\\\textbf{Key-based PLS:} Mutual information, key generation rate.}} \\
  \cline{1-2}
  Mobile RIS & RIS on pedestrians/vehicles/ships & ~ \\
  \cline{1-2}
  Flying RIS &  RIS on UAVs/airships/satellites & ~ \\
  \hline
\end{tabular}
\end{table*}
However, most of the considered scenarios focus on RIS-aided keyless PLS; hence, the research on RIS-aided key-based PLS is still relatively scarce. The combination of PLKG with RISs entails new challenges and opportunities. On one hand, the control of the RIS can change the time-varying characteristics of wireless fading channel, which increases the randomness of the channel entropy source. Moreover, the increase of randomness does not introduce additional noise to the channel features. In this context, we begin with several potential application scenarios of RIS-aided PLKG, which are shown in Fig. \ref{Fig:RIS_scenario}. The first scenario is the basic indoor RIS-aided PLKG, where Eve attempts to eavesdrop the channel information and recover the raw key. The second scenario considers a multi-RIS cooperation that is used to further increase the bit generation rate and reduce the information leakage on Eve via the cooperative RIS control strategy. In the third scenario, the direct link between Alice and Bob is blocked, and a RIS can be deployed at the obstacle edge to boost the PLKG. The optimization deployment of RIS is critical for facilitating key generation and preventing information leakage. The fourth application scenario shows that given the wireless coverage enhancement assisted by the RIS, Eve can hide in the corner of the obstacle to eavesdrop the key information, which increases the covertness of Eve. The fifth scenario illustrates that the RIS can be utilized by a malicious user to launch denial of service (DoS) attacks to further prevent key generation and communication. The last three are attack-defense scenarios, where both the legitimate users and malicious user are equipped with RIS, and the direct link is blocked by an obstacle. Thus, the control strategy of the RIS between the white hat and hacker is an interesting research direction. Furthermore, RIS-aided PLKG among multi-user and/or multi-antenna are also potential application scenarios, which can further increase the key generation rate.

Despite the variety of application scenarios, there are some design challenges worth considering. First, whether the presence of a RIS can break the channel reciprocity required for key generation. Second, whether a RIS can improve the key generation rate by increasing the randomness of the channel entropy source without limit. Third, where is the optimal RIS deployment location for RIS-aided PLKG and is the law of optimal deployment the same as that in RIS-assisted wireless communications. Last but not the least, what is the influence of RIS-enabled physical layer attack and how to mitigate this attack. Next, we articulate the design challenges in RIS-aided PLKG with the experimental case study from a legal user and an attacker perspective, respectively.

\section{Challenges in RIS for PLKG Scheme}
\begin{figure*}[!ht]
 \centering
  \includegraphics[width=14.5cm]{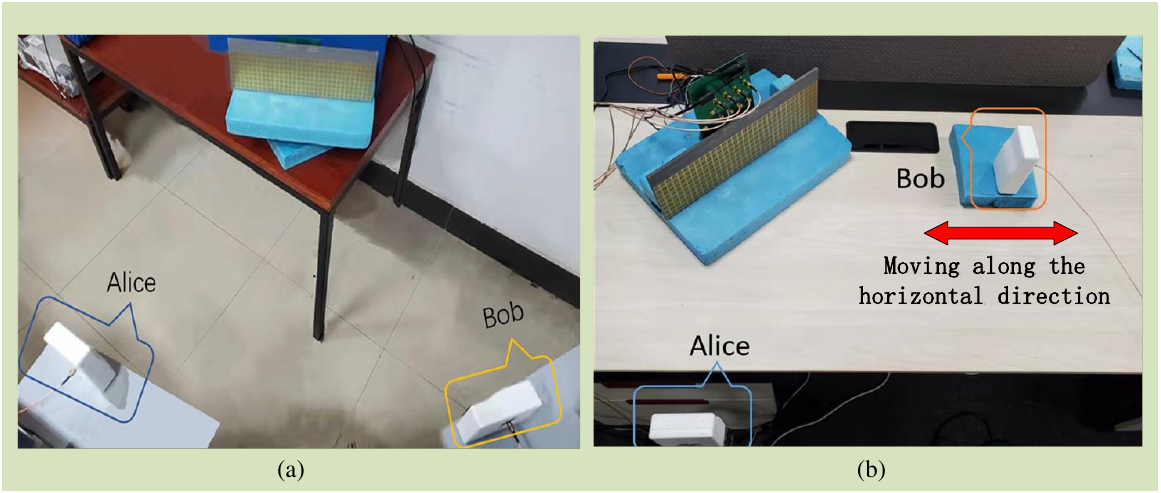}
  \caption{The experiment representation for RIS-aided PLKG. Figure 2(a) is for the experiment of channel
reciprocity and RIS reconfiguration speed, where Alice, Bob and RIS are at fixed positions. Figure 2(b) is for the experiment of RIS deployment, where Alice and RIS are static, while Bob can move along the horizontal direction.}
  \label{Fig:RIS_deployment}
\end{figure*}
In this section, we analyze the main challenges in designing RIS-aided PLKG from proof-of-concept experiments, which includes channel reciprocity, RIS reconfiguration speed and RIS deployment. The experiment scenario is shown in Fig. \ref{Fig:RIS_deployment}; therein, the channel reciprocity and RIS reconfiguration speed are based on Fig. \ref{Fig:RIS_deployment}(a), and the experiment of RIS deployment is based on the scenario in Fig. \ref{Fig:RIS_deployment}(b).

Regarding the experiment architecture, the experiment requires the usage of the high-performance notebook HOST PC, the software radio platform USRP-RIO, the synchronous clock node WR LEN, the clock distributor WR switch, and the RIS. The RIS operates at 4.25 GHz, and has 8 rows $\times$ 32 columns RIS units, where each RIS unit is a square with size 0.012 m and every two rows of the RIS units are a group. The RIS is controlled by four voltage signals V1, V2, V3, V4, where each voltage signal can randomly generate one of 16 equal interval voltages within $[0,21]$ to control four groups of RIS units. Alice and Bob are configured with a single antenna and connected to the PC terminal via an optical fiber. Regarding the signal design, each wireless frame is 10 ms and 2 OFDM symbols with 1,200 sub-carriers for each symbol are inserted to each frame. The first OFDM symbol is inserted with a pilot at every 6 sub-carriers, and a total of 200 pilots are inserted for the acquisition of CSI. The second OFDM symbol includes 1,200 bit of data for the calculation of RSS. The least squares (LS) channel estimation is used to obtain the CSI, and the double threshold quantization method and single bit cumulative distribution function (CDF) quantization method are adopted to obtain the raw bit sequences \cite{CDFquantization}. All the raw key sequences are carried on randomness tests based on the statistical test suite for random number generators (NIST).

\subsection{Channel Reciprocity}
For channel reciprocity, we discuss whether the addition of RIS can preserve the channel reciprocity or not.
\subsubsection{Setup}
For this experiment setup, the Alice-Bob transmitter pair is placed in a static indoor environment, where channel fading is varying slowly and there are few scatterers. Alice and Bob are at fixed locations and the distance between Alice and Bob is about 1.5 m. In order to compare it to the case without RIS, we first collect 20,000 frames in the Alice-Bob pair and refer to as DATA 1. Next, by setting the RIS reconfiguration speed to be 100 ms/time, we extract one frame every ten frames and record 20,000 frames on the PC terminal which is referred as DATA 2.
\subsubsection{Results \& Analysis} With CSI estimation, RSS calculation and bit quantization for DATA 1 and DATA 2, respectively, we obtain the raw bit sequences. From the results in Fig. \ref{Fig:PGR11}(a), we find that the bit mismatch rate using CSI is significantly lower than the bit mismatch rate using RSS, both in double threshold quantization and single bit CDF quantization. The reason is that RSS represents the average signal energy over a period of time and is a coarse-grained measurement of channel information. On the contrary, CSI is a kind of fine-grained channel information measurement, which can better reflect the characteristics of the channel. Most obviously, from the results of DATA 1 and DATA 2 in Fig. \ref{Fig:PGR11}(a), we see that the bit mismatch rate of DATA 1 is significantly higher than that of
DATA 2, which indicates that the received signal energy can be enhanced with the assistance of the RIS in each reflected path. In other words, the additive Gaussian white noise is one of the factors affecting channel reciprocity. The transceiver can improve the channel reciprocity by increasing the signal to noise ratio (SNR), and then reduce the bit mismatch rate, which facilitates the raw
key generation. The CDF of RSS for DATA 1 and DATA 2 can be found in Fig. \ref{Fig:PGR11}(b); the CDF of DATA 1 changes rapidly with RSS, whereas this change is slower in DATA 2. This observation showcases that with the addition of RIS, the RSS values have a wider temporal fluctuation range, which can effectively increase the channel randomness. Notably, Fig. \ref{Fig:PGR11}(d) showcases that the key generation rate yields a 15-fold improvement, which underlines the potential of RIS in PLKG.
\subsection{RIS Reconfiguration Speed}
In this experiment, by analyzing different RIS reconfiguration speeds, we reveal the influence of RIS control on PLKG.
\subsubsection{Setup} In this experiment, Alice and Bob are also placed in a static indoor environment with fixed locations. Apart from changing the RIS reconfiguration speed with 100 ms/time, just as in the channel reciprocity experiment, we adjust the RIS reconfiguration speed with 1,000 ms/time and collect one frame in every 100 frames. After recording 20,000 frames on the PC terminal, we denote the data set as DATA 3.
\subsubsection{Results \& Analysis} 
\begin{figure*}[!ht]
 \centering
  \includegraphics[width=16.3cm]{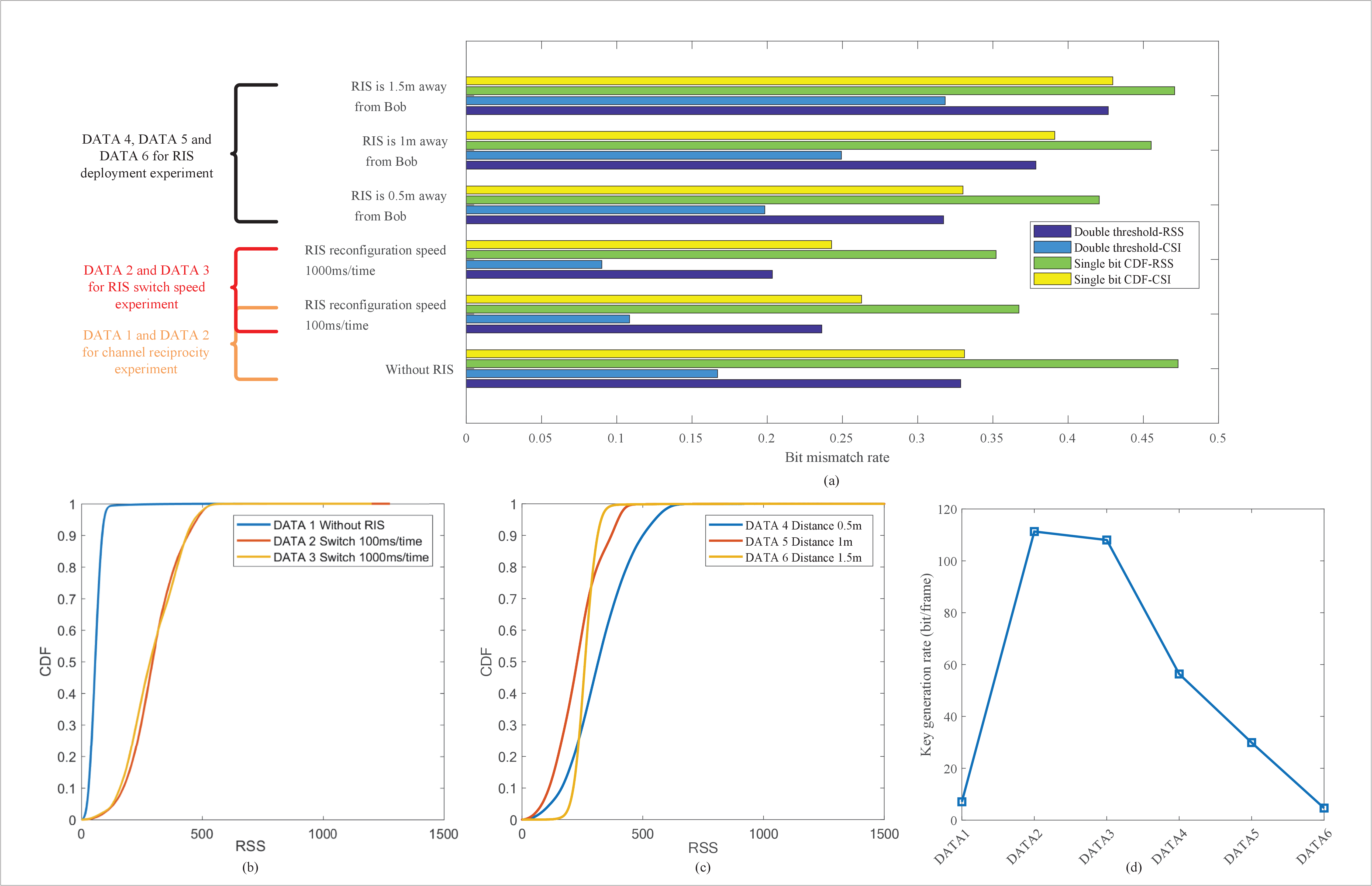}
  \caption{The experiment results of RIS-aided PLKG design.}
  \label{Fig:PGR11}
\end{figure*}

DATA 3 is also transformed into raw bit sequences via CSI estimation, RSS calculation and bit quantization. In Fig. \ref{Fig:PGR11}(a), compared with DATA 1 and DATA 2, the bit mismatch rate of DATA 3 is the smallest. The bit mismatch rate of DATA 3 is smaller than DATA 1 that is owing to the channel reciprocity enhancement of RIS. Compared the bit mismatch rate of DATA 2 with DATA 3, it suggests that the RIS reconfiguration speed is proportional to the bit mismatch rate. A very fast RIS reconfiguration has a negative impact on the bit match due to the channel probing time will greater than the channel coherence time. Even so, from Fig. \ref{Fig:PGR11}(d), we find that the key generation rate of DATA 3 is slightly lower than that of DATA 2. This partly thanks to the information reconciliation process of PLKG and partly due to the successful quantization of raw bits form the fluctuating channel. The experimental results clearly prove the challenge of optimizing the RIS reconfiguration speed to attain a tradeoff between key generation rate and bit mismatch rate. In addition, from Fig. \ref{Fig:PGR11}(b), we can further conclude that the RIS plays a constructive role in PLKG with appropriate designs.
\subsection{RIS Deployment}
The deployment of RIS gives a new degree of freedom for PLKG, which creates new challenges and opportunities. Here, we reveal the impact of RIS deployment on PLKG.
\subsubsection{Setup}
In a static indoor scenario, Alice's location is fixed and 0.5 m away from the RIS, while Bob can move horizontally to different locations, thereby representing different RIS deployments. Specifically, the RIS reconfiguration speed is 100 ms/time, thus one frame is extracted and stored from every ten frames. We collect three data sets, DATA 4, DATA 5, and DATA 6 with respect to the distance between RIS and Bob at 0.5 m, 1 m and 1.5 m, respectively.
\subsubsection{Results \& Analysis} In the same way, with the obtained CSI via LS, the double threshold quantization and single bit CDF quantization are utilized to produce the raw bits. The results in Fig. \ref{Fig:PGR11}(a) show that when the distance between Bob and RIS grows, the bit mismatch rate gets higher, which indicates the following two things. On one hand, the longer the distance between two communication participants, the greater the path loss will be. Thus, the low SNR results in poor channel quality, which leads to a higher bit mismatch rate. On the other hand, the longer the distance between Bob and RIS is, the proportion of the energy of the RIS reflected path in the energy of all scattered paths will decrease, and then the RIS will no longer play a major role in controlling the wireless environments. As a result, with the channel reciprocity decreasing, the bit mismatch rate increases. This can also be seen from Fig. \ref{Fig:PGR11}(d) where the key generation rate of DATA 6 decreases significantly or even becomes smaller than that without RIS. Thus, these observations suggest that the RIS deployment and the communication distance of transceiver are important factors for a high key generation rate. Furthermore, this experiment also proves that CSI has a better performance than that of RSS for PLKG with different RIS deployments. Moreover, from Fig. \ref{Fig:PGR11}(c), we can observe that the closer RIS is to the transceiver, the larger the channel fluctuation becomes. Thus, the study of optimization deployment of RIS for PLKG is a potentially interesting direction.
\section{RIS for Attack Scheme: Feasibility and Countermeasures}
In this section, we study the case that the RIS is controlled by Eve and give some insightful results from experiments. Then, we discuss challenges and possible countermeasures.
\subsection{Challenges \& Experiments}
When malicious Eve has control over a RIS, it can rapidly flip the RIS to change the wireless environments, which can affect the PLKG between Alice and Bob. Here, this attack is referred to as RIS jamming attack. Next, we show the influence of such an attack by analyzing the RIS reconfiguration speed during PLKG as a case.
\subsubsection{Setup}
In this scenario, the location of Alice is fixed and the distances between Bob and RIS are 0.5 m, 1 m and 1.5 m, respectively. Particularly, the RIS is assumed to be controlled by Eve and the RIS reconfiguration speed is set to be 1 ms/time. According to the different distances between Bob and RIS, we collect three data sets, namely, DATA 7, DATA 8 and DATA 9, respectively.
\begin{figure}[!ht]
 \centering
  \includegraphics[width=7.6cm]{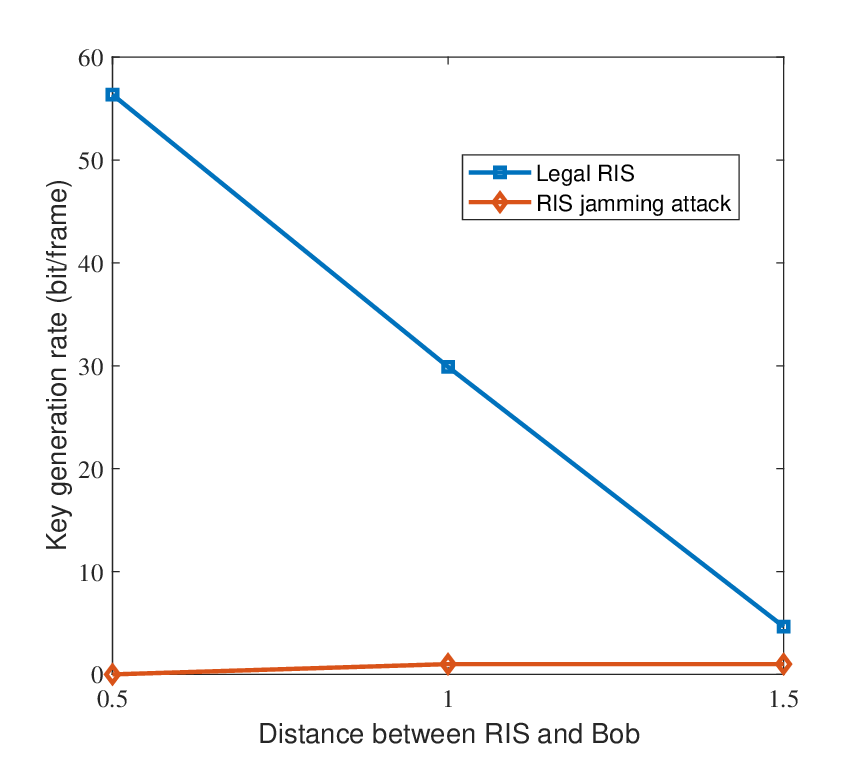}
  \caption{The key generation rate: Legitimate RIS vs. RIS jamming attack.}
  \label{Fig:RISat}
\end{figure}
\subsubsection{Results \& Analysis}
Based on the raw bit sequences generated from DATA 7, DATA 8 and DATA 9, we observe that at the same distance between Bob and RIS, the bit mismatch rate with RIS reconfiguration speed 1 ms/time is much higher than that of 100 ms/time. The reason is that when the RIS reconfiguration speed is changed to 1 ms/time, the transceiver pairs cannot complete at least one frame bidirectional channel probing within a coherence time. That is, the channel reciprocity between Alice and Bob is seriously compromised. This attack produces a great pressure on information reconciliation for PLKG, which means higher negotiation overhead. Fig. \ref{Fig:RISat} shows that when the RIS is reconfigured at 1 ms/time by Eve, the key generation rate in different distances is only 0$\sim$1 bit/frame, which seriously hinders the PLKG process of the legitimate transceivers.
\subsection{Discussions}
The RIS jamming attack is an active attack which can change the RIS reconfiguration speed in a very short time, i.e., in the order of $\mu s$, which can be regarded as the channel coherence time being compressed by configuring $\mathbf{\Phi}$. In this case, Alice and Bob cannot perform bidirectional channel probing, effectively. Thus, the RIS attack is a DoS attack, which restrains the normal process of PLKG. Unfortunately, different from traditional active attack countermeasures, which can trace such attack based on wireless radio frequency (RF) fingerprints, RIS jamming attack is inherently undetectable due to the passive nature of RIS which has no RF chain. Therefore, some new ideas are necessary to defend against such detrimental attacks.
\begin{itemize}
  \item \textbf{Countermeasure 1}: Since this attack may achieve its malicious objective by changing the wireless environment using a RIS, similarly, legitimate users can deploy a RIS to counteract the changes of radio electromagnetic wave, thus extending the coherence time of the channel. Specifically, Alice and Bob can superimpose a multiplicative channel, i.e., $\mathbf{h}^T_{RB}\mathbf{\Phi}\mathbf{h}_{AR}$, to counteract the influence of the illegal multiplicative channel, i.e., $\mathbf{h}^T_{EB}\mathbf{\Phi}_E\mathbf{h}_{AE}$, on the wireless signal. The challenge of this defense strategy is that the legitimate users need to know the RIS reconfiguration speed model of attack. Deep learning and reinforcement learning are powerful tools to learn the control strategy of a RIS in a model-free approach, thereby confronting the attack strategy. Furthermore, multi-RIS can be used by legitimate users to simultaneously resist this attack and improve the PLKG rate in a static indoor environment.
  \item \textbf{Countermeasure 2}: The RIS jamming attack induces a rapid variation of the wireless signal's amplitude and phase, which makes bidirectional channel probing fail. In this case, we can choose more robust features of the wireless channel to generate the raw key sequences. Specifically, although the bidirectional channel probing for RSS or CSI is different in the presence of RIS jamming attack, the legitimate users share the physical space and the scatterers. In this case, we can deploy multiple antennas on the legitimate users and extract the long-time-scale CSI features to generate the raw key. For example, the delay and angle of each path are bidirectional reciprocal and can be captured based on Newton orthogonal matching pursuit (NOMP) algorithm. Then, we can robustly generate a key without considering the damage caused by the RIS jamming attack. Since the spatial reciprocity features are often slow-varying, we have to deploy RISs for legitimate users to artificially create a time-varying random scatter environment or design high resolution multi-bit quantization method to increase the key generation rate.
  \item \textbf{Countermeasure 3}: From the experiment of channel reciprocity, we see that the assistance of RIS improves the SNR of PLKG. Inspired by this, we can detect the RIS attack by monitoring the variation of RSS. Specifically, we have to obtain the statistical property of RSS in both secure and attack cases. Then, a binary hypothesis test can be used to decide whether there is a RIS jamming attack in current wireless security situation. When an attack occurs, the Alice-Bob pair stops the PLKG process to reduce the unnecessary resource consumption. However, this countermeasure is a passive defense strategy which can detect the existence of RIS jamming attack but cannot eliminate such attack to achieve the fundamental PLKG purposes. Therefore, to transform a passive defense strategy to an active defense strategy, we can use the integrated sensing and communications (ISAC) or a fine-grained wireless link signature to trace back the physical location of the attack and mitigate it.
\end{itemize}

\section{Conclusions and Future Directions}

In this article, we have provided an overview of the promising RIS-aided PLKG technology in a static indoor environment. Notably, a RIS increases the SNR, thereby improving the channel reciprocity. The RIS reconfiguration speed cannot increase without limit, there is a tradeoff between key generation rate and bit mismatch rate, but with a suitable RIS reconfiguration speed, the overall key generation rate is 15-fold higher than that without RIS. Next, we find that it is better to deploy the RIS near the transceiver side to reinforce the function of RIS in controlling the electromagnetic waves. We also highlight that the optimizations of RIS reconfiguration speed and deployment are meaningful directions. Furthermore, the investigation of RIS attack-defense for RIS-aided PLKG is a valuable direction. We further identify several open challenges for RIS-aided PLKG in the following.
\begin{itemize}
  \item \textbf{RIS and/or multi-RIS cooperative configuration in near/far field RIS-aided PLKG during communication}: The channel models of near field and far field are different. When the multipath fading and interference are insufficiency in near/far field, there is a strong correlation between the legitimate channel and wiretap channel even if their protected zone is more than half a wavelength away. One possible direction is the RIS and/or multi-RIS optimization control, thereby obtaining an achievable key secrecy rate. Interestingly, by configuring the RIS, the transceiver can realize dual-functional integration that is improving PLKG while communicating. This gives birth to a promising direction of 6G endogenous security, that we propose and name as integrating communications and security (ICAS), where the dual functions of communication and security can mutually benefit by sharing spectrum, power and hardware, etc.
  \item \textbf{RIS-aided PLKG in frequency division duplex (FDD) systems}: Different from TDD systems, FDD systems lack reciprocity on the frequency related channel parameters between uplink and downlink. One possible solution is to control RIS and/or multi-RIS to artificially create a large number of frequency independent reciprocal channel parameters to generate the secret key, i.e., angle of arrival and/or departure, etc. Besides, for massive MIMO FDD systems, the RIS-aided PLKG can be combined with CSI feedback, which can reconstruct the reciprocal channel.
  \item \textbf{Defense strategy of PLKG for RIS attacks}: RIS attacks can realize an intelligent and programmable jamming, eavesdropping and spoofing. Then, by injecting the malicious reflection parameters to the multiplicative channel, it can leak the secret key information and cause a series of threats in PLKG process. Particularly, given the passive components of RIS, RIS attacks are often highly covert. On these lines, the countermeasures for RIS attacks in PLKG are still largely open.
  \item \textbf{RIS-aided PLKG in ubiquitous 6G wireless networks}:
       The RIS-aided PLKG can be used in ubiquitous wireless networks, ranging from autonomous vehicles, wearable electronics to drones and deep space satellites. It provides new degrees of freedom to configure RIS-aided PLKG to support new IoE services, thereby improving the security of communication, sensing and computing, etc.
\end{itemize}

\ifCLASSOPTIONcaptionsoff
  \newpage
\fi



\bibliographystyle{IEEEtran}
\bibliography{IEEEabrv,sigproc} 

\begin{thebibliography}{10}
\providecommand{\url}[1]{#1}
\csname url@samestyle\endcsname
\providecommand{\newblock}{\relax}
\providecommand{\bibinfo}[2]{#2}
\providecommand{\BIBentrySTDinterwordspacing}{\spaceskip=0pt\relax}
\providecommand{\BIBentryALTinterwordstretchfactor}{4}
\providecommand{\BIBentryALTinterwordspacing}{\spaceskip=\fontdimen2\font plus
\BIBentryALTinterwordstretchfactor\fontdimen3\font minus
  \fontdimen4\font\relax}
\providecommand{\BIBforeignlanguage}[2]{{%
\expandafter\ifx\csname l@#1\endcsname\relax
\typeout{** WARNING: IEEEtran.bst: No hyphenation pattern has been}%
\typeout{** loaded for the language `#1'. Using the pattern for}%
\typeout{** the default language instead.}%
\else
\language=\csname l@#1\endcsname
\fi
#2}}
\providecommand{\BIBdecl}{\relax}
\BIBdecl

\bibitem{7539590}
Y.~Liu, H.-H. Chen, and L.~Wang, ``Physical layer security for next generation
  wireless networks: Theories, technologies, and challenges,'' \emph{IEEE
  Commun. Surveys Tuts.}, vol.~19, no.~1, pp. 347--376, Jan. 2017.

\bibitem{8883129}
L.~Jiao, N.~Wang, P.~Wang, A.~Alipour-Fanid, J.~Tang, and K.~Zeng, ``Physical
  layer key generation in {5G} wireless networks,'' \emph{IEEE Wireless
  Commun.}, vol.~26, no.~5, pp. 48--54, May 2019.

\bibitem{6311224}
Q.~Wang, K.~Xu, and K.~Ren, ``Cooperative secret key generation from phase
  estimation in narrowband fading channels,'' \emph{IEEE J. Sel. Areas
  Commun.}, vol.~30, no.~9, pp. 1666--1674, Sep. 2012.

\bibitem{9000831}
N.~Aldaghri and H.~Mahdavifar, ``Physical layer secret key generation in static
  environments,'' \emph{IEEE Trans. Inf. Forensics Security}, vol.~15, pp.
  2692--2705, Feb. 2020.

\bibitem{1528749}
T.~Aono, K.~Higuchi, T.~Ohira, B.~Komiyama, and H.~Sasaoka, ``Wireless secret
  key generation exploiting reactance-domain scalar response of multipath
  fading channels,'' \emph{IEEE Trans. Antennas Propag.}, vol.~53, no.~11, pp.
  3776--3784, Nov. 2005.

\bibitem{9326394}
Q.~Wu, S.~Zhang, B.~Zheng, C.~You, and R.~Zhang, ``Intelligent reflecting
  surface-aided wireless communications: A tutorial,'' \emph{IEEE Trans.
  Commun.}, vol.~69, no.~5, pp. 3313--3351, May 2021.

\bibitem{LIASKOS20191}
C.~Liaskos~\textit{et al}, ``A novel communication paradigm for high capacity
  and security via programmable indoor wireless environments in next generation
  wireless systems,'' \emph{Ad Hoc Networks}, vol.~87, pp. 1--16, May 2019.

\bibitem{9298937}
Z.~Ji~\textit{et al}, ``Secret key generation for intelligent reflecting
  surface assisted wireless communication networks,'' \emph{IEEE Trans. Veh.
  Technol.}, vol.~70, no.~1, pp. 1030--1034, Jan. 2021.

\bibitem{9569556}
P.~Staat~\textit{et al}, ``Intelligent reflecting surface-assisted wireless key
  generation for low-entropy environments,'' in \emph{Proc. IEEE PIMRC}, Sep.
  2021, pp. 745--751.

\bibitem{9771319}
G.~Li~\textit{et al}, ``Reconfigurable intelligent surface for physical layer
  key generation: Constructive or destructive?'' \emph{IEEE Wireless Commun.},
  vol.~29, no.~4, pp. 1--8, Aug. 2022.

\bibitem{8957701}
Z.~Ji~\textit{et al}, ``Vulnerabilities of physical layer secret key generation
  against environment reconstruction based attacks,'' \emph{IEEE Wireless
  Commun. Lett.}, vol.~9, no.~5, pp. 693--697, May 2020.

\bibitem{CuiTieJunRIS}
T.~J. Cui~\textit{et al}, ``Coding metamaterials, digital metamaterials and
  programmable metamaterials,'' \emph{Light: Science \& Applications}, vol.~3,
  no. e218, Oct. 2014.

\bibitem{9656117}
X.~Pang, N.~Zhao, J.~Tang, C.~Wu, D.~Niyato, and K.-K. Wong, ``{IRS}-assisted
  secure {UAV} transmission via joint trajectory and beamforming design,''
  \emph{IEEE Trans. Commun.}, vol.~70, no.~2, pp. 1140--1152, Feb. 2022.

\bibitem{9198898}
A.~Almohamad, A.~M. Tahir, A.~Al-Kababji, H.~M. Furqan, T.~Khattab, M.~O.
  Hasna, and H.~Arslan, ``Smart and secure wireless communications via
  reflecting intelligent surfaces: A short survey,'' \emph{IEEE Open J. Commun.
  Soc.}, vol.~1, pp. 1442--1456, 2020.

\bibitem{CDFquantization}
N.~Patwari, J.~Croft, S.~Jana, and S.~Kasera, ``High-rate uncorrelated bit
  extraction for shared secret key generation from channel measurements,''
  \emph{IEEE Trans. Mobile Comput.}, vol.~9, no.~1, pp. 17--30, Jan. 2010.

\end{thebibliography}
%

%
%








\end{document}